\documentclass[11pt]{article}
\usepackage[left=1.5cm,right=1.5cm]{geometry}
\usepackage[numbers,square]{natbib}
\usepackage[T1]{fontenc}
\usepackage{tikz-cd}
\usepackage[british]{babel}
\usepackage{layout}
\usepackage{amsmath}
\usepackage{amssymb,amstext, amsthm, simplewick, amsfonts}
\usepackage{framed}
\usepackage{amsfonts}
\usepackage{color} 
\usepackage{braket}
\usepackage{multicol} 
\usepackage{epsfig}
\usepackage{cancel}
\usepackage{caption}
\usepackage{hyperref}
\usepackage{epsf}
\usepackage{feynmf} 
\usepackage{frontespizio}
\usepackage{rotating}
\usepackage{subfigure}
\usepackage{pstricks}
\usepackage{type1ec}
\usepackage{lettrine}
\usepackage{bbold}
\usepackage{calligra}
\usepackage{tikz}
\usepackage{float}
\usepackage{subfigure}
\usepackage{slashed}
\usepackage{mathrsfs}
\usepackage{epstopdf}
\usepackage{curve2e}
\usepackage{setspace}
\usepackage{indentfirst}
\usepackage{emptypage}
\usepackage{relsize}
\usepackage{mathrsfs}
\usepackage{graphicx}
\usepackage{stackengine}
\usepackage{calc}
\usepackage{hyperref}
\hypersetup{
	colorlinks,
	citecolor=green,
	filecolor=black,
	linkcolor=blue,
	urlcolor=black
}
\def\sqb{\,\raise.5pt\hbox{$\overline{\nbox{.09}{.09}}$}\,}
\newcommand{\3}[1]{C_{
		\ifthenelse{\equal{\ThreePt}{\empty}}{#1}{
			\ifthenelse{\equal{#1}{\empty}}{\ThreePt}{\ThreePt,#1}}}}
\newcommand{\redef}[1]{{C'}_{
		\ifthenelse{\equal{\ThreePt}{\empty}}{#1}{
			\ifthenelse{\equal{#1}{\empty}}{\ThreePt}{\ThreePt,#1}}}}
\newcommand{\ren}[1]{C_{
		\ifthenelse{\equal{\ThreePt}{\empty}}{#1}{
			\ifthenelse{\equal{#1}{\empty}}{\ThreePt}{\ThreePt,#1}}}}
\newcommand{\sd}[1]{D_{
		\ifthenelse{\equal{\ThreePt}{\empty}}{#1}{
			\ifthenelse{\equal{#1}{\empty}}{\ThreePt}{\ThreePt,#1}}}}

\numberwithin{equation}{section} 
\makeatletter
\@addtoreset{equation}{section}
\newcommand{\bea}{\begin{eqnarray}}
\newcommand{\eea}{\end{eqnarray}}
\makeatother
\newcommand{\beqa}{\begin{eqnarray}}
	\newcommand{\eeqa}{\end{eqnarray}}

\newcommand{\sm}{\mathcal{S}}
\newcommand{\nn}{\nonumber}

\let\i=\iota        \let\m=\mu
\let\n=\nu

\newcommand{\bann}{\begin{eqnarray*}}
\newcommand{\eann}{\end{eqnarray*}}

\def\nbox#1#2{\vcenter{\hrule \hbox{\vrule height#2in
			\kern#1in \vrule} \hrule}}
\def\sq{\,\raise.5pt\hbox{$\nbox{.09}{.09}$}\,}
\def\sqb{\,\raise.5pt\hbox{$\overline{\nbox{.09}{.09}}$}\,}
\def\Box{\sq}
\usepackage{accents}

\newcommand{\bmi}{\begin{minipage}}
\newcommand{\emi}{\end{minipage}}

  \let\D=\Delta

\newcommand{\be}{\begin{equation}}
	\newcommand{\ee}{\end{equation}}
\newcommand{\sdfrac}[2]{\mbox{\small$\displaystyle\frac{#1}{#2}$}}

\newcommand{\beq}{\begin{equation}}
	\newcommand{\eeq}{\end{equation}}

\newcommand{\ThreePt}{\empty}
\usepackage{accents}
\makeatletter
\newcommand{\xLine}[2][]{\ext@arrow 0359\Rightarrowfill@{#1}{#2}}
\makeatother
\xdefinecolor{darkgreen}{RGB}{102, 204, 70}
\xdefinecolor{darkblue}{RGB}{0, 0, 153}
\usepackage{amssymb}
\usepackage{pifont}

\usepackage{tikz-feynman}
\usetikzlibrary{arrows}
\usetikzlibrary{snakes}
\usetikzlibrary{decorations,decorations.pathmorphing,decorations.markings}

\tikzset{graviton/.style={decorate, decoration={snake}, double}}
\tikzset{gluon/.style={decorate, decoration={coil, segment length=8, aspect=1.2, amplitude=3 }}}

\begin{document}
	\begin{center}
		\vspace{1.5cm}
	\begin{center}
	\vspace{1.5cm}
	{\Large \bf Conformal Backreaction, Chiral and Conformal Anomalies \\ }
\vspace{0.3cm}
	{\Large \bf in the Early Universe}\footnote{Presented by Claudio Corian\`o at the XVII Marcel Grossmann Meeting, Pescara, Italy, 7-12 July 2024} 
	 
\vspace{0.3cm}
	
		\vspace{1cm}
	{\bf Claudio Corian\`o,  Stefano Lionetti, Dario Melle, Riccardo Tommasi\\}
	{ and\\}
	{\bf Leonardo Torcellini\\}
		\vspace{1cm}
{\it  Dipartimento di Matematica e Fisica, Universit\`{a} del Salento \\
and INFN Sezione di Lecce,Via Arnesano 73100 Lecce, Italy\\
National Center for HPC, Big Data and Quantum Computing\\}
\vspace{0.5cm}

	\end{center}
	\begin{abstract}
The backreaction of a conformal matter sector and its associated conformal anomaly on gravity can be systematically studied using the formalism of the anomaly effective action. This action, defined precisely in flat spacetime within ordinary quantum field theory, can be analyzed perturbatively in terms of external graviton insertions. The expansion coefficients correspond to correlation functions of the stress-energy tensor, which are renormalized through two key counterterms: the square of the Weyl tensor $(C^2)$ and the Gauss-Bonnet term $(E)$. Anomalous conformal Ward identities impose hierarchical constraints on this expansion, revealing that the anomaly's contribution arises from bilinear mixings of the form $R \Box^{-1} E$ and $R \Box^{-1} C^2$, supplemented by local Weyl-invariant terms. These mixings reflect the non-local structure of the anomaly. The precise form of the effective action, however, may vary depending on the regularization scheme used, with potential differences manifesting through additional Weyl-invariant terms. 
These actions encapsulate the breaking of Weyl invariance in the early universe, with implications that are particularly relevant during the inflationary epoch. For chiral and gravitational anomalies, we demonstrate that the corresponding effective actions exhibit similar structures, influencing the evolution of chiral asymmetries in the early universe plasma.
 
\end{abstract}
	\end{center}
	\newpage
	
	\section{Introduction}
	
General Relativity (GR) can be extended or modified by incorporating higher-order terms involving the Riemann tensor and its contractions, or by introducing additional fields beyond the metric tensor, such as scalar fields, leading to various forms of dilaton gravity models. Through field redefinitions, with the inclusion of Lagrange multipliers, and Weyl transformations at intermediate stages, these actions can be reformulated into different forms that lend themselves to specific interpretations. What uniquely defines the most convenient form of the action is the predicted spectrum of the quantum excitations, with their definitive mass eigenstates. \\
Such modifications are often proposed to address phenomena not fully explained by classical GR, especially in extreme conditions like high-energy regimes or strong gravitational fields. There is a prevailing notion that the Einstein-Hilbert (EH) action, even when supplemented by a cosmological constant ($\Lambda$) to account for the standard cosmological \(\Lambda\) cold dark matter 
($\Lambda$CDM) model, may require corrections as we approach the Planck scale. In these high-curvature regimes, significant deviations from traditional GR could arise, potentially offering an explanation for the nature of dark energy, rendering it dynamical.\\
For instance, simpler modifications might involve quadratic corrections to GR. These modifications, along with higher-order corrections, can be derived from functional integration over a matter sector, in line with Sakharov's concept of induced gravity. In this framework, in fact, even the gravitational (EH) action emerges from the evaluation of quantum effects due to ordinary matter fields, (see 
\cite{Visser:2002ew} for a review). \\
Additionally, higher-order derivative terms in the gravitational action are being investigated, particularly in the context of higher-derivative gravity theories, with motivations of the pursuit of a renormalizable theory of gravity. \\
Another significant modification involves $f(R)$ gravity, where the EH action is generalized to a function of the Ricci scalar. The simplest modification, known as the Starobinsky model, enlarges the EH Lagrangian ($\sim \sqrt{g} R$) with the inclusion of quadratic corrections, and has been widely studied for its success in explaining the inflationary expansion of the early universe 
\begin{equation}
\mathcal{S} =  M_P^2 \int d^4x \sqrt{-g} \, \left[ R + \alpha R^2 \right] + \mathcal{S}^{(m)}.
\label{prima}
\end{equation}
In \eqref{prima} \( \mathcal{S}^{(m)} \) represents the action of matter fields and $M_P$ is the Planck mass. The \( R^2 \) term generates a mechanism for cosmic inflation, the phase of accelerated expansion in the early universe (see for example \cite{FGaus,Kehagias:2013mya}).
While these extensions offer new ways to address challenges in GR, they also present their own difficulties. \\Quadratic corrections, for example, must be consistent with empirical observations, such as solar system tests and gravitational wave detections, while also adhering to theoretical principles like causality and energy conditions. The modifications induced are substantial.\\
 In vacuum (i.e., when \(T_{\mu\nu} = 0\)), the field equations indicate that the Ricci scalar \(R\) is no longer constant but instead evolves according to a higher-order differential equation. The dynamics of \(R\) play a pivotal role in determining the behavior of the solutions to these equations. In the case of \( R^2 \) gravity, as described by equation \eqref{prima}, the action can be reformulated as an effective scalar-tensor theory, where the additional degree of freedom introduced by the \( R^2 \) term is interpreted as a scalar field coupled to gravity. This scalar field, often denoted as \(\varphi\), takes on the role of the inflaton in the Starobinsky inflationary model.

What makes \( R^2 \) gravity particularly compelling is that it provides a natural mechanism for cosmic inflation in the early universe, eliminating the need to artificially introduce an external scalar field.  \section{Issues with modifications of GR}
A major theoretical concern in higher-derivative gravity theories is the issue of unitarity and the potential emergence of ghosts—unphysical degrees of freedom associated with instabilities in the theory. In some specific cases, such as Lovelock theory \cite{Lovelock:1971yv}, carefully constructed higher-curvature terms avoid the ghost problem, ensuring a consistent theory. However, in general, the introduction of higher-order terms increases the complexity of the theory, making it challenging to preserve unitarity and avoid ghosts unless the coefficients of these terms are chosen with great care. \\
Among the quadratic corrections, particular attention is given to the Gauss-Bonnet term
\begin{align}
\label{fourd}
E_4&\equiv R_{\mu\nu\alpha\beta}R^{\mu\nu\alpha\beta}-4R_{\mu\nu}R^{\mu\nu}+R^2  
\end{align}
which arises naturally in higher-dimensional gravity theories. In four dimensions, this term is a topological invariant, but its inclusion in the context of the gravitational effective action requires a careful analysis of the renormalization procedure, which is not unique. \\
The analysis of this term in the context of the renormalization of conformal matter sectors coupled to gravity is only briefly reviewed in this talk, while additional details can be found in \cite{Coriano:2022ftl}.
In general, its inclusion in a gravitational theory plays a key role in theories with extra dimensions, making it an important focus of study. In certain modifications of the Einstein-Hilbert action, such as the one we are going to consider, the term is also accompanied by the square of the Weyl tensor 
\beq
 ( C^{(4)})^2\equiv R_{\mu\nu\alpha\beta}R^{\mu\nu\alpha\beta}-2R_{\mu\nu}R^{\mu\nu}+\frac{1}{3}R^2. 
 \eeq
 The terms \(E\) (the Euler density) and \(C^2\) (the square of the Weyl tensor) naturally arise due to the anomalous breaking of conformal symmetry and manifest in the trace of the renormalized stress-energy tensor of matter. They  will be the focus of our analysis, as we explore the impact of conformal backreaction from conformal matter on gravity. The aim of this investigation is to identify the so-called conformal anomaly effective action, which can be added the Einstein-Hilbert (EH) action or any other gravitational metric-based action to study the cosmology of the early universe.\\
The key assumption underlying this approach is that conformal symmetry may have played a fundamental role in the early universe, particularly in the high-energy regimes, central to cosmology and particle physics. This symmetry is a defining feature of several foundational theories, including quantum field theory and string theory, both of which suggest that the universe, in its earliest moments at extremely high energies, could have exhibited conformal symmetry. Understanding the conformal anomaly that signals the breaking of conformal symmetry and its backreaction on gravity, may provide critical insights into the physics of the early universe and the mechanisms that shaped its evolution.
 \subsection{More general anomalies of even and odd parity}
 The impact of anomalies in the early universe doesn't stop to the conformal phase. Fermionic and bosonic (for spin-1) chiral asymmetries can be generated and enhanced by the chiral anomaly in the primordial plasma.  
 At the quantum level, the fermionic chiral symmetry experiences a profound violation, which reveals itself through the non-conservation of the axial current 
 \beq
 \label{five}
  J^\mu_{5 f} = \bar{\psi} \gamma^\mu \gamma_5 \psi . 
\eeq
This violation, in the presence of both Abelian and gravitational backgrounds with field-strength $F_{\mu\nu}$ and Riemann tensor $R^\alpha_{\, \, \beta \mu\nu}$, is summarized by the following covariant anomaly equation
\beq
\label{rr}
\nabla_\mu J^\mu_{5 f} = a_1 \varepsilon^{\mu\nu\rho\sigma} F_{\mu\nu}F_{\rho\sigma} + a_2 \varepsilon^{\mu\nu\rho\sigma} R^\alpha_{\, \, \beta \mu\nu}R^\beta_{\, \,\alpha \rho\sigma}
\eeq
where \( a_1 \) and \( a_2 \) are constants. Associated with this current is a chiral charge measuring the difference between the number of left-handed and right-handed modes of the fermion sector. Similar chiral asymmetries can be induced in the bosonic case, when $J_{5}$ is realized by a Chern-Simons current \cite{Dolgov:1987yp,delRio:2020cmv} (see the discussion in \cite{Coriano:2023gxa}).
First identified within high-energy physics, the chiral anomaly’s ramifications extend far beyond this realm. Its influence is profound, touching on topics from cosmology \cite{Kamada:2022nyt} to condensed matter physics \cite{Chernodub:2021nff}, with notable examples including the quantum Hall effect, the chiral magnetic effect, and applications involving topological insulators \cite{Coriano:2024ive}.\\
Long ago it was proposed that chiral (parity-odd) and conformal (parity-even) anomalies could be considered together at quantum-level, through the deviation of the energy-momentum tensor from being traceless \cite{Deser:1980kc}. Its general expression takes the form
\beq
\label{ee}
g_{\mu \nu}\left\langle T^{\mu \nu}\right\rangle = b_1 E + b_2 C^{\mu \nu \rho \sigma} C_{\mu \nu \rho \sigma} + b_3 \nabla^2 R + b_4 F^{\mu \nu} F_{\mu \nu} + f_1 \varepsilon^{\mu \nu \rho \sigma} R_{\alpha \beta \mu \nu} R_{\rho \sigma}^{\alpha \beta} + f_2 \varepsilon^{\mu \nu \rho \sigma} F_{\mu \nu} F_{\rho \sigma}.
\eeq
Here, the constants \( b_1, b_2, b_3, b_4, f_1, \) and \( f_2 \) govern different contributions to the anomaly. 
$b_1$ and $b_2$ parameterize the parity-even part, while $b_3$ identifies a term which is of even parity but prescription 
dependent. The remaining coefficients multiply the parity-odd terms of a trace anomaly.  An interesting debate has ensued recently concerning the actual generation of such terms in the context of 
a Lagrangian field theory, with opposite conclusions. We refer to  \cite{Bonora:2014qla,Armillis:2010pa,Bastianelli:2018osv,Abdallah:2021eii,Abdallah:2023cdw,Larue:2023tmu} for more details. In a recent study on non-Lagrangian conformal field theories, it was demonstrated that such anomalies cannot be excluded a priori \cite{Coriano:2023cvf}. These anomalies, whether chiral or conformal, highlight essential deviations from classical symmetries and have a profound influence on the physics of the early universe.

\section{Simple estimates }
One crucial aspect of our considerations involves a possible conformal symmetric phase of our universe and its breaking.
To characterize such phase phase we can adopt a simplified approximation by assuming a single phase transition occurred in the early universe, roughly at the electroweak scale temperature, \( T_{\text{EW}} \sim 245 \, \text{GeV} \). At this temperature, the Hubble parameter can be estimated to reflect the dynamics of this transition, providing a clearer picture of the breaking of conformal symmetry, given by the relation
\beq
H(T) = \sqrt{\frac{8 \pi^3 g_*(T)}{90}} \frac{T^2}{M_{\text{Pl}}},
\eeq
where \( g_*(T) \) is the effective number of relativistic degrees of freedom at temperature \( T \), equal approximately to 100, and corresponding to a time $
t \sim {1}/{2H}$, assuming radiation dominance. $M_{\text{Pl}} \approx 2.435 \times 10^{18} \, \text{GeV}$ is the reduced Planck mass. This corresponds to $H(T_{\text{EW}}) \approx 4.1 \times 10^{-5} \, \text{GeV}$, and a time $t_{\text{EW}} \sim 8 \times 10^{-21} \, \text{seconds}$.\\
 A similar back-of-the-envelope calculation gives, for a scale of inflation $(\sim T_{inf}\sim 10^{15})$ a time $t\sim 10^{-38}$ seconds, which usually ends at 
$t\sim 10^{-30} s$, when reheating starts, setting up the conditions for the hot Big Bang.\\ The universe transitions to the radiation-dominated era. Within our assumptions, this time interval defines the era where conformal symmetry is present, together with the anomaly contribution. \\
The conformal anomaly, as described by \eqref{ee} adds corrections to the Hubble parameter $H$
 which governs the expansion rate of the universe during inflation. 
\subsection{Parity-even terms}
In the conformal phase, we focus on the standard case of the parity-even terms in \eqref{ee}, which involve the Euler density, \(E\), and the Weyl tensor squared, \(C^2\).\\
The Weyl tensor \(C_{\mu\nu\lambda\rho}\) measures the conformally invariant part of the spacetime curvature. In a flat de Sitter space, which is conformally flat, the Weyl tensor vanishes, \(C_{\mu\nu\lambda\rho} = 0\), and so does the \(C^2\) term. However, in realistic inflationary scenarios where spacetime is not perfectly de Sitter (due to perturbations, gravitational waves, or anisotropies), the Weyl tensor becomes non-zero, and \(C^2\) scales with the deviations from de Sitter symmetry. If we relate the curvature at a certain stage of the early universe to the Hubble parameter, noticing that  
$R\sim H^2$ in an inflationary phase, then one may expect that the anomaly contribution is of $O(H^4)$. \\
The size of the conformal anomaly’s effect on inflation depends on the relative magnitude of the anomaly-induced energy density compared to the inflaton’s potential energy. The energy density from the anomaly is typically given by
\beq
\label{crt}
\rho_{\text{anomaly}} \sim \frac{H^4}{(4\pi)^2} \sum_i N_i,
\eeq
where \( H \) is the Hubble parameter during inflation and \( N_i \) is the number of degrees of freedom of the quantum fields. If \( H \sim 10^{12} \, \text{GeV} \) during inflation (as in many high-scale inflation models), and if \( N_i \sim 100 \) (for Standard Model particles and beyond), we estimate:
\beq
\rho_{\text{anomaly}} \sim \frac{(10^{12})^4}{(4\pi)^2} \times 100 \sim 10^{46} \, \text{GeV}^4.
\eeq
For comparison, the typical energy density during inflation is
\beq
\rho_{\text{inf}} \sim 3 M_{\text{Pl}}^2 H^2 \sim 10^{43} \, \text{GeV}^4.
\eeq
The ratio of the anomaly energy density to the inflaton energy density is extraordinarily large
\beq
\frac{\rho_{\text{anomaly}}}{\rho_{\text{inf}}} \sim \frac{10^{46}}{10^{43}} \sim 10^3.
\eeq
This indicates that the anomaly could be quite significant at the inflationary scale, potentially contributing corrections to the inflationary dynamics. However, the exact impact depends on the specific inflation model, the quantum fields involved, and their interaction with the anomaly.\\
The situation is different at current time.
To quantify the impact of the conformal anomaly on dark energy, we can follow a similar approach to how we estimate the effect during inflation. The energy density contribution from the anomaly is still proportional to \( H^4 \), where \( H \) is the Hubble parameter. In the present-day universe, where \( H_0 \sim 10^{-33} \, \text{eV} \), the energy density due to the anomaly is again given by \eqref{crt}, with $H$ replaced by $H_0$.\\
Using the current estimate of $H_0$, and assuming a similar number of degrees of freedom as in the Standard Model we can infer that $\rho_{\text{anomaly}} \sim 10^{-133} \, \text{eV}^4.$ This value is extremely small compared to the observed dark energy density, which is on the order of \( 10^{-11} \, \text{eV}^4 \). \\
Thus, while the conformal anomaly contributes a small correction to dark energy, it might not be large enough to account for the entire dark energy density. However, in specific models, especially those involving interactions with quantum fields or curvature, the anomaly's contribution could be enhanced, leading to observable effects.
There are several unresolved issues associated with these estimates, particularly regarding the role of the anomaly during the inflationary epoch. This concern arises from the nature of the Gauss-Bonnet term, which is topological unless it is coupled with a scalar field—a scenario that is certainly plausible.\\
Despite these open questions that warrant careful consideration, we can still draw some general conclusions about the impact of anomaly corrections on current cosmological models. It is reasonable to expect that the conformal anomaly directly influences the tensor perturbations (gravitational waves) produced during inflation, making them a subject worthy of further investigation.

\section{The role of conformal symmetry and its breaking}
Conformal symmetry can be broken both by an anomaly and/or spontaneously.\\
 The breaking of this symmetry probably defines one of the most difficult issues in the physics of the fundamental interactions since the underlying Lagrangian describing conformal matter is not allowed to contain any scale. In general this prolem is solved 
 by introducing a dilaton field in the Lagrangian that acquires a vacuum expectation values (vev) and triggers the spontanous breaking of the symmetry. The central issue  is how the dilaton field in a conformal theory can take a vev if the Lagrangian contains no explicit scale. \\
 This is indeed a subtle problem, as introducing an explicit mass or scale would break the conformal symmetry explicitly, rather than allowing for spontaneous symmetry breaking.
However, at the quantum level, radiative corrections can modify the potential. These corrections can generate a dimensional transmutation, whereby a dimensionless coupling (like \(\lambda\)) acquires a logarithmic dependence on the renormalization scale \(\mu\). For instance, for a scale-invariant potential of the form 
$V(\phi)=\lambda \phi^4$, this leads to a one-loop effective potential of the form

\beq
V_{\text{eff}}(\phi) = \lambda(\mu) \phi^4 + \frac{\beta(\lambda)}{16\pi^2} \phi^4 \ln\left(\frac{\phi^2}{\mu^2}\right),
\eeq
where \(\beta(\lambda)\) is the beta function for the coupling \(\lambda\). This logarithmic correction can generate a non-trivial minimum for the potential, at which the dilaton field \(\phi\) acquires a vev, thus breaking the conformal symmetry spontaneously.\\
The conformal anomaly has  a similar effect, being related to the renormalization of the theory in the ultraviolet.
The following phase of cosmological inflation, with the stretching of spacetime that ensues, may have left an imprint of such combined breakings on current cosmological observables.  \\
In this talk, we focus solely on the anomalous breaking of conformal symmetry, following two primary lines of inquiry. First, we explore the reconstruction of the anomaly effective action in flat space, guided by constraints from conformal symmetry and the standard perturbative expansion. Second, we examine the anomaly-induced action, which aims to capture the breaking of symmetry for any background metric, using a functional solution to the anomaly constraint \eqref{ee} in the parity-even case.\\
Notably, these two approaches do not fully converge beyond a certain order in gravitational fluctuations. Systematic tests of these relationships for well-defined correlators have been conducted up to 4-point  with agreement observed up to 3-point functions \cite{Coriano:2017mux}, after which discrepancies arise \cite{Coriano:2022jkn}.

 \section{The effective action from the conformal backreaction: perturbative tests}
To illustrate the mechanism of conformal backreaction,  we consider a matter sector coupled to gravity. For simplicity we consider the case of a scalar field $\chi$, conformally coupled at $d=4$.\\
The backreaction of a conformal sector on the gravitational metric can be analyzed through the partition function $\mathcal{Z}_B(g)$, defined in the Euclidean metric as

\beq
\mathcal{Z}_B(g) = \mathcal{N}\int D\chi \, e^{-S_0(g,\chi)},
\eeq
where $\mathcal{N}$ is a normalization constant. The effective action, $\mathcal{S}_B(g)$, is given by

\beq
e^{-\mathcal{S}_B(g)} = \mathcal{Z}_B(g) \quad \text{or} \quad \mathcal{S}_B(g) = -\log \mathcal{Z}_B(g).
\eeq

This effective action collects multiple insertions of the stress-energy tensor. 
Diagrammatically, $\mathcal{S}_B(g)$ is expressed in terms of stress-energy tensor correlators, which can be computed via a Feynman expansion

\begin{align}
\label{figgx1}
\sm(g)=& \sum_n \quad\raisebox{-8.5ex}{{\includegraphics[width=0.19\linewidth]{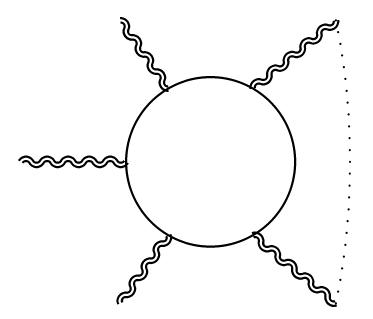}}} \,\scriptstyle \text{(n-point)}
\end{align}

For a scalar field $\chi$, the stress-energy tensor is defined in terms of the classical action $\sm_0$ as

\beq
T^{\mu\nu}_{\text{scalar}} = \frac{2}{\sqrt{g}}\frac{\delta S_0}{\delta g_{\mu\nu}} = \nabla^\mu \chi \, \nabla^\nu \chi - \frac{1}{2}g^{\mu\nu}g^{\alpha\beta} \nabla_\alpha \chi \, \nabla_\beta \chi + \chi \left(g^{\mu\nu}\Box - \nabla^\mu \nabla^\nu + \frac{1}{2}g^{\mu\nu}R - R^{\mu\nu} \right) \chi^2,
\eeq
which is inserted $n$ times, perturbatively, in the bare loop of \eqref{figgx1}.
In a flat background, this expansion occurs order by order in $1/M_P^2$, accounting for metric fluctuations $h_{\mu\nu}$.\\
Its bare quantum average is defines as usual as 
\beq
\langle T^{\mu\nu}\rangle_B \equiv \frac{2}{\sqrt{g}} \frac{\delta \mathcal{Z}_B}{\delta g_{\mu\nu}}.
\eeq
 These contributions are divergent as $d \to 4$ and require renormalization. Explicit computations in dimensional regularization (DR) show that  the ultraviolet divergences of the effective action  can be removed just by adding two counterterms, $V_E$ and $V_{C^2}$,  defined as integrals over the Euler density and Weyl tensor squared, respectively
\beq
V_{C^2}(g, d) = \mu^{\epsilon} \int d^dx \sqrt{-g} \, C^2, \quad V_E(g, d) = \mu^{\epsilon} \int d^dx \sqrt{-g} \, E,
\eeq
where $\mu$ is the renormalization scale and $\epsilon = d - 4$. 
Their inclusion breaks the conformal symmetry in the effective action since in DR

\begin{align}\label{finalvar}
2g_{\mu \nu}\frac{\delta}{\delta g_{\mu \nu}}\int d^d x  \sqrt{g} (C^{(4)})^2&=
(d-4)\sqrt{g}\left[(C^{(4)})^2+\frac{2}{3}\Box R \right],
\end{align}
and 
\begin{equation}
2g_{\mu \nu}\frac{\delta}{\delta g_{\mu \nu}}\int d^d x  \sqrt{-g} E=(d-4)\sqrt{g}  E.
\end{equation}
In the case of a general matter content, where scalars are accompanied by fermions and spin-1 fields in the virtual corrections, we can define the regularized effective action in the form $(\epsilon=d-4)$
\begin{equation}
\label{rena}
\mathcal{S}_R(g,d)=\lim_{\epsilon\to 0}\left( \mathcal{S}_B(g,d) +  b' \frac{1}{\epsilon}V_E(g,d) + b \frac{1}{\epsilon}V_{C^2}(g,d)\right),
\end{equation}
where $b$ and $b'$ are related to the conformal matter content.

\subsection{The expansion around flat space}
In practice, the simplest way to investigate the anomaly effective action is through a standard perturbative expansion around flat space, with renormalization performed as described in \eqref{rena}. One can utilize the conformal Ward identities (CWIs) to test the consistency of hierarchies satisfied by correlators with, say, $n$ external gravitons in terms of correlators with $n-1$, $n-2$, and so on gravitons. The effective action around flat space includes all terms—both Weyl-invariant and anomalous (i.e., Weyl-variant)—while maintaining diffeomorphism invariance. \\
The result of this analysis, presented in \cite{Coriano:2021nvn}, shows that the Weyl-variant terms, which respect the anomaly constraints, are characterized by insertions of contributions like $R\Box^{-1}$ on the external graviton legs. This combination defines a dimensionless expansion parameter, where $R$ is the linearized Ricci scalar, and $\Box^{-1}$ represents the Green function of the ordinary D'Alembertian in flat space. The Weyl-variant hierarchy, as decomposed in \cite{Coriano:2021nvn}, is consistently structured with the inclusion of an additional term, called the "zero-residue" term, which ensures the anomaly hierarchy is fully self-consistent.\\
In other words it satisfies all the symmetries mentioned above.  Details can be found in the original work. A similar pattern emerges from the analysis of another 4-point function in \cite{Coriano:2022jkn}. The anomaly manifests in a graviton-scalar mixing analogous to the spin-1-spin-0 mixing emerging from the expansion of a gauge theory around a spontaneously broken vacuum. We recall that expanding a gauge theory around a nontrivial vacuum $(v)$, for example in the electroweak sector, generates the term  $g\, v \,\partial_\mu \, G_0\, Z^{\mu}$ where $G_0$ is the Goldstone mode of the broken gauge symmetry and $Z^{\mu}$ denotes the spin-1 gauge boson, for example the neutral $Z$ gauge boson. In the case of the anomaly action constructed from perturbation theory, the effective interaction is summarized by nonlocal bilinear terms of the form $R\Box^{-1} E$ and $R\Box^{-1} C^2$, with the operators $E$ and $C^2$ accounting for the correct mass dimension. 
\subsection{Additional symmetries and the extraction of the correlators of gravitons}
The action can be computed quite efficiently up to the 4th or 5th order in free field theories.
Additional symmetries constrain the correlation functions starting from the renormalized one-point function $\braket{T^{\mu\nu}(x)}$. This satisfies the fundamental Ward identity of covariant conservation in an arbitrary background $g$
\begin{equation}
\label{vat}
\nabla_\mu\braket{T^{\mu\nu}(x)}=0, \qquad i.e. \qquad \delta_\epsilon \sm_R=0
\end{equation}
as a consequence of the invariance of $\sm_R(g)$ under diffeomorphisms $x^\mu\to x^\mu +\epsilon^\mu(x).$
Here $\nabla_\mu$ denotes the covariant derivative in the general background metric $g_{\mu\nu}(x)$. It can be expressed in the form 
\begin{equation}
\partial_\nu\left(\frac{\delta \sm_R(g)}{\delta g_{\mu\nu}(x)}\right)+\Gamma^\mu_{\nu\lambda}\left(\frac{\delta \sm_R(g)}{\delta g_{\lambda\nu}(x)}\right)=0\label{conserv},
\end{equation}
where $\Gamma^\mu_{\lambda\nu}$ is the Christoffel connection for the general background metric $g_{\mu\nu(x)}$. Functional differentiations of \eqref{vat} allow to derive constrain on correlators of higher orders which are hierarchical. 
The stress-energy correlators for fluctuations around a background metric, say $\bar{g}$, can then be combined into a functional $\mathcal{S}_B(g)$ that collects all the correlation functions containing multi-graviton vertices 
as coefficients of the expansion in the gravitational perturbations around flat space

\beq
\mathcal{S}_B(g) = \mathcal{S}(\bar{g}) + \sum_{n=1}^\infty \frac{1}{2^n n!} \int d^dx_1 \cdots d^dx_n \sqrt{g_1} \cdots \sqrt{g_n} \langle T_1 \cdots T_n \rangle_{\bar{g}} \delta g_1 \cdots \delta g_n,
\eeq
represented by the expansion in \eqref{figgx1}. They are defined as 
\begin{equation}
\label{exps1}
\langle T^{\mu_1\nu_1}(x_1)\ldots T^{\mu_n\nu_n}(x_n)\rangle_B \equiv\frac{2}{\sqrt{g_1}}\ldots \frac{2}{\sqrt{g_n}}\frac{\delta^n \sm_B(g)}{\delta g_{\mu_1\nu_1}(x_1)\delta g_{\mu_2\nu_2}(x_2)\ldots \delta g_{\mu_n\nu_n}(x_n)}, 
\end{equation}
where $\sqrt{g_1}\equiv \sqrt{|\textrm{det} \, g_{{\mu_1 \nu_1}}(x_1)} $ and so on, constrained by Ward identities and anomalous CWIs that take specific hierarchical forms \cite{Coriano:2017mux}. In the comparison with the perturbative approach, the metric $\bar{g}$ is chosen to be the ordinary flat metric of Minkowski spacetime.
A similar expansion can be generated by replacing the bare action $\sm_B$ with its renormalized version $\sm_R$.
These expansions have been investigated in several previous works in free field theories realizations up to fourth order in the gravitational fluctuations, where they provide the most direct way to understand their structure \cite{Coriano:2021nvn,Coriano:2018bsy,Coriano:2023lmc}. \\
The renormalization of this functional expansion is rather involved in a general background \cite{Coriano:2022ftl}, and can be best understood by borrowing from the DR prescription around flat space. \\
The perturbative analysis offers crucial insights into the structure of the effective action, enabling a meaningful comparison with the formal derivation of the anomaly-induced action. In the latter approach, the effective action is derived through a variational solution of the anomaly constraint \eqref{ee} in the parity-even case. Specifically, in \(d=4\), this method combines a Weyl transformation with the elimination of the conformalon field, which controls the metric rescaling. The outcome is a nonlocal action, which we will briefly outline below.
We will examine these contributions as they arise from two perspectives: first, from the general solution to the anomaly constraint, represented by the nonlocal anomaly action; and second, from the analysis of scalar, fermion, and vector theories around flat space, using free-field theory realizations of the interactions. This dual approach allows us to explore how conformal backreaction on gravity is influenced by the exchange of conformal sectors in quantum corrections.

\begin{figure}[t]
\label{figgx}
 \centering
	\raisebox{-1.5ex}{\includegraphics[scale=0.5]{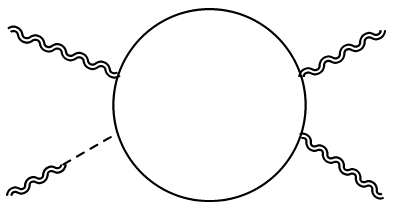}} 
	\raisebox{-1.5ex}{\includegraphics[scale=0.5]{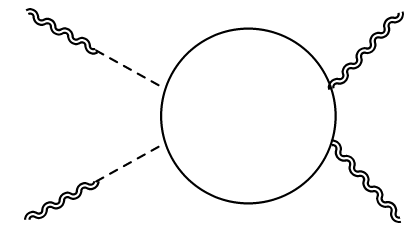}} 
	\raisebox{-1.5ex}{\includegraphics[scale=0.5]{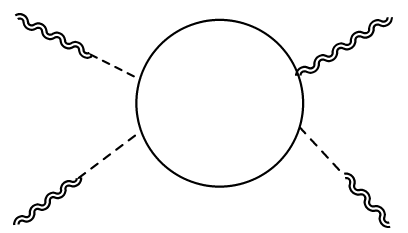}} 
	\raisebox{-1.5ex}{\includegraphics[scale=0.5]{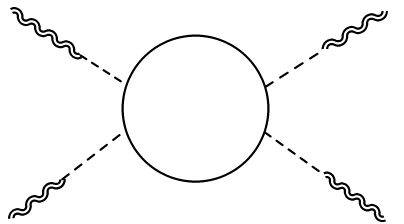}} 
	\caption{The Weyl-variant contributions from $\mathcal{S}_A$  to the renormalized vertex for the 4T with the corresponding bilinear mixings in $d=4$ \label{4T}, in the decomposition of \cite{Coriano:2022ftl}.}
\end{figure}

\section{Expanding the counterterms for general backgrounds}
The expansion of the counterterms \eqref{finalvar} is a critical step that needs a very close attention, 
while from $\mathcal{S}_B(g,d)$, the bare effective action, we extract its singularities in the $d\to$ 4 limit \cite{Coriano:2022ftl}. Indeed, at $d=4$, the integration of a conformal sector induces a renormalized effective action $\sm_R$ in \eqref{rena}, whose variation under an infinitesimal Weyl transformation of the metric 
 \beq
\label{vars}
g_{\mu\nu}\to e^{2 \sigma(x)} g_{\mu\nu},\qquad 
\delta_\sigma g_{\mu\nu}= 2 \sigma g_{\mu\nu} 
\eeq
 is equal to the conformal anomaly. \\
 The identity
 \beq
 \frac{\delta}{\delta \sigma}F[g]=2 g_{\mu\nu}\frac{\delta}{\delta g_{\mu\nu}}F[g]
 \eeq
shows that trace of the energy momentum tensor induced by a certain (Weyl invariant) functional of the metric, say $F[g]$, 
is related to its behaviour under \eqref{vars}. Therefore, the variation \eqref{vars} on the functional $\sm_R(g)$ gives the relation involving the stress energy tensor $T^{\mu\nu}$
\begin{equation}
\label{anomx}
\frac{\delta \sm_R}{\delta \sigma(x)}=\sqrt{g} \,g_{\mu\nu}\,\langle T^{\mu\nu}\rangle, 
\qquad \langle T^{\mu\nu}\rangle = \frac{2}{\sqrt{g}}\frac{\delta \sm_R}{\delta g_{\mu\nu}}.
\end{equation} 
 There are ambiguities in the extraction of the finite part of the loop corrections in $\mathcal{S}_B$, a procedure that is not uniquely defined in curved spacetime. For instance, a formal expansion of $\mathcal{S}_B$ can be organised in the form  
\begin{equation}
\label{pone}
\sm_B(g,d)= \lim_{d\to 4}\left(\sm_f(d) -\frac{b}{\epsilon}V_{C^2}(g,4) - \frac{b'}{\epsilon}V_E(g,4)\right),
\end{equation}
where $\sm_f$ is finite. \\
Notice that \(V_E(g,4)\) is purely topological, and its definition highlights the challenges in establishing a consistent formulation. This difficulty arises because it depends on the specific regularization prescription used. Such issues have been encountered in the study of Einstein-Gauss-Bonnet (EGB) gravities, where one attempts to bypass Lovelock's theorem by performing a \(0/0\) limit of the Gauss-Bonnet term. This is achieved through an infinite renormalization of the coupling constant in front of the Gauss-Bonnet term, effectively making it dynamical, much like what is done for the Einstein-Hilbert action in \(d = 2\).\\
Eq. \eqref{pone} is justified by the fact that a conformal sector generates only singularities with  a single $1/\epsilon$ pole in all the correlators, and that these can be canceled just by the inclusion of $V_{C^2}$, accompanied by the evanescent term $V_E$.  
 Explicitly, under a small local variation of the metric \eqref{vars} one derives the general expression
\begin{equation}
 \delta_\sigma \sm_R=\frac{1}{(4 \pi)^2}\int d^4 x \sqrt{g}\delta\, \sigma(x)\left(c_1 R_{\mu\nu\rho\tau}R^{\mu\nu\rho\tau} + c_2 R_{\mu\nu}R^{\mu\nu} +c_3 R^2 + c_4\square R\right), 
\label{var1}
\end{equation}
which is constrained by the Wess-Zumino consistency condition 
\begin{equation}
\label{WZ}
\left[\delta_{\sigma_1},\delta_{\sigma_2}\right]\sm_R=0,
\end{equation}
for two independent variations $\delta \sigma_1$ and $\delta {\sigma_2}$.
 It can be shown that the coefficients $c_i$ have to satisfy the relation $c_1+c_2 +3 c_3=0$, allowing to re-express \eqref{var1} in the form 
 \begin{equation}
 \label{anof}
 \delta_\sigma\sm_R=\frac{1}{(4 \pi)^2} \int d^4 x \sqrt{g}\delta\sigma(x) \mathcal{A}(x)
 \end{equation}
 where
\begin{equation}
 \label{vv}
 \sqrt{g} \,g_{\mu\nu}\,\langle T^{\mu\nu}\rangle=\mathcal{A}(x)=\left( a E + b C^2 + c\Box R\right) 
\end{equation}
is the parity-even conformal anomaly.
The coefficients $a,b,c$ are automatically fixed by the conformal sector that is integrated out, and the contributions $E$ and $C^2$ are both part of the variation of the renormalized effective action, generated by $V_E$ and $V_{C^2}$ contained in \eqref{rena}. \\
The expression of the effective action is then linked to the regularization procedure implemented on the  bare action.  In general, this procedure generates effective actions which take the form of dilaton gravities, where the conformal factor introduced in the decomposition of the metric 

\beq
 g_{\mu\nu}=\bar{g}_{\mu\nu}e^{2 \phi}
\label{dec}
\eeq
appears in the expansion of the counterterms around $d=4$ by the relation 
\begin{equation} 
\label{expand1}
\frac{1}{\varepsilon}V_{E/C^2}(g,d)=\frac{\mu^{\epsilon}}{\epsilon}\left( V_{E/C^2}(g,4) + \epsilon 
V_{E/C^2}'(g,4) +O(\varepsilon^2) \right), \qquad \epsilon \to 0,
\eeq
where 
\beq
\label{oone1}
V'_{E/C^2}=\lim_{\epsilon \to 0}\frac{1}{\epsilon}\left( V_{E/C^2}(g,d)-V_{E/C^2}(g,4)\right).
\eeq
The renormalized effective action 
then takes the form 
\begin{equation}
\label{simp}
\sm_R\equiv \sm_R(4)=\sm_f (4)  + V'_E(g,4) + V'_{C^2}(g,4)
\end{equation}
expressed in terms of a finite, Weyl invariant contribution $\sm_f (4)$ and the anomaly part, related to the 
$V'_{E/C^2}(g,4)$ terms introduced by the renormalization procedure. \\
The anomaly is generated, after renormalization, by the non-invariance of $\sm_R\equiv \sm_R(4)$ under a Weyl redefinition of the metric 
\beq
\sm_R[\bar {g}e^{2 \phi}]\neq  \sm_R[\bar {g}].
\eeq
This implies that the invariance of the effective action under Weyl transformations is broken, and the choice of a fiducial metric \(\bar{g}\) becomes relevant. Hence, \(\bar{g}\) cannot be chosen arbitrarily. In other words, the invariance under the conformal decomposition \eqref{dec} is broken by the renormalization procedure, and from the original metric \(g_{\mu\nu}\), we uniquely identify a conformalon/dilaton field \(\phi\).\\
\subsection{Renormalizations: DR and WZ subtractions}
Since the approach discussed above may appear too formal, it is convenient to discuss it in simple physical terms. 
The dilaton $\phi$ in the parameterizaton \eqref{dec} carries no physical dimensions. In order to canonically identify as a physical fields of mass dimension one, we need to redefine it in the form 
$\phi=\bar{\phi}/f$, where the redefinition depends on the space time point. We can consider a spacetime  region where $f$ can be taken to be constant. The field $\phi$ fluctuates around the value $f$ which essentially takes the role of its vacuum expectation value.  
Notice that prior to any breaking of the Weyl symmetry, in each local spacetime region a free-falling observer 
 would verify the presence of a conformal symmetry in tangent space which is broken by the renormalization procedure. This is natural since Weyl invariance in a curved background has to generate a $SO(2,4)$ symmetry in any tangent plane. The effective action that one identifies in a local patch is characterised by the vev of the dilaton field in that patch. Notice that the breaking of the conformal symmetry is induced by the radiative corrections.   \\ 
The expansion \eqref{expand1}, has been denoted as "DR subtraction" in \cite{Coriano:2022ftl}, since extends to a curved spacetime the usual DR approach from Minkowski space. 
A second possibility is given by the functional expansion
\begin{align}
\label{rdef}
&\hat{V}'_E( g, \phi)=\lim_{\epsilon\to 0}\frac{1}{\epsilon}\left(V_E(g,d)-V_E(\bar g,d)\right)
\end{align}
which differs from the previous regularization by Weyl invariant terms. In this case the subtractions in the bare action $\sm_B$ are defined with respect to the fiducial metric with the decomposition
\beqa
\sm_R(d)&=&\Big( \sm_B(g,d)  +\frac{1}{\epsilon}V_{C^2}(\bar g,d) +\frac{1}{\epsilon}V_E(\bar g,d)\Big) +
\frac{1}{\epsilon} \left( V_{E}(g,d) 
- V_{E}(\bar g,d)  \right) + 
\frac{1}{\epsilon} \left( V_{C^2}(g,d) 
-V_{C^2}(\bar g,d)  \right), \nn
\eeqa
in the $d\to 4$ limit,
with the Weyl invariant, finite contributions given by the expression
\beq
\label{sf2}
\tilde\sm_f(4)=\lim_{d\to 4} \Big(\sm_B(d)  +\frac{1}{\epsilon}V_{C^2}(\bar g,d) +\frac{1}{\epsilon}V_E(\bar g,d)\Big).
\eeq
The renormalized effective action then takes the form 
\beqa
\label{sumxp}
\sm_R(4)&=&\tilde\sm_f(4) +\sm_{WZ},
\eeqa
where
\beq
\label{WZ}
\sm_{WZ}\equiv \hat{V}'_E(\bar g,\phi)+ \hat{V}'_{C^2}(\bar g,\phi).
\eeq
This specific definition of the counterterms, expanded in their dependence around $d=4$ and expressed in terms of the full metric and of the fiducial metric, as clear from \eqref{WZ}, is commonly used in the derivation of the  Wess-Zumino form (WZ) of the anomaly action.\\
It can be shown that the use of the regularization in the form given above, by subtracting $V_E(\bar g,d)$ in 
$d$ dimensions - rather than at $d=4$ - is eliminating some Weyl invariant terms \cite{Coriano:2022ftl}.
In summary, a consistent procedure can be established to extract the effective action at \(d=4\) from the singular limit of a topological term. This approach is carried out in \(d\) dimensions and involves some intermediate technical steps, for which details can be found in \cite{Coriano:2022ftl}.   
The approach has been used in more recent studies of classical dilaton gravities incorporating only the $0/0$ limit of the Gauss-Bonnet tems (also known as 4d EGB theories) \cite{Coriano:2022ftl}.
They identify two forms of the EGB theory, here indicated as $EGB_1$ and $EGB_2$
\beqa
 \label{sxc}
 \sm_{EGB/1}&=&S_{EH} + V'_E(\bar{g},\phi) \nn \\
 \sm_{EGB/2}&=&S_{EH} + \hat V'_E(\bar{g},\phi), 
 \eeqa
 The term $0/0$ refers to the fact that $E$ is topological and its contribution to the equations of motion is therefore metric independent and vanishes at $d=4$. If the dimensionless cooupling $1/g_E$ is also rescaled to infinity then one can obtain a finite contribution to the action coming from the topological term. This results in a form of dilaton gravity that differs substantially from the WZ form of the action  
In both formulations, the EH action in \eqref{sxc} is re-expressed in terms of the fiducial metric and the dilaton field. In the case of conformal anomaly actions, the addition of the quantum corrections identified as $\sm_f$ and 
 $\tilde{\sm}_f$ as well of $V'_E(\bar{g},\phi)$ and $\hat{V}'_E(\bar{g},\phi)$ to \eqref{sxc} allows to quantify the conformal backreaction on the external classical background.  \\
 If we follow the regularization of \eqref{pone}, \eqref{simp}, the effective action takes the form 
 
\begin{align} \label{EGBW1}
&\tilde S_{EGBW_1} = \frac{1}{16\pi G}\int d^4 x \sqrt{g} \  e^{-2\phi} \left[ R + 6\nabla_\lambda \phi \nabla^\lambda \phi ] -2e^{-2\phi}\Lambda\right]  + \mathcal{S}_f(4) \nonumber \\
&+\int d^4 x \sqrt{g} \biggl[-\phi(b' E+ b C^2)-b'\left(4 G^{\mu\nu}(\nabla_\mu\phi\nabla_\nu\phi)+2(\nabla_\lambda \phi\nabla^\lambda \phi)^2-4\bar\Box\phi\nabla_\lambda \phi\nabla^\lambda \phi\right)\biggl],
\end{align}
where the first term on the right hand side is generated by the gauging of the EH action with the inclusion of a cosmological constant $\Lambda$. The second term  $\mathcal{S}_f(4)$, identifies the contribution from the renormalized loop corrections extracted from the bare action $\sm_B$ as in \eqref{pone}, and the third 
term is the contribution from the anomaly, represented by the $V'_{E/C^2}(\bar{g},\phi)$ terms. 
Notice that in this regularization, the anomaly couples through terms of the form $\phi C^2$ and $\phi E$. In particular, the Gauss-Bonnet term, which is topological in nature, loses its purely topological character due to its interaction with the dilaton field $\phi$. A final comment concerns the pure gravitational action, that in 
\eqref{EGBW1} is generated from the EH action by a Weyl decomposition of the metric. One could equally consider a replacement of the EH action by a conformal invariant action
\beq
\label{cf}
\sm=\int d^4 x \sqrt{g} C^2.
\eeq
Also in this case, a conformal decomposition as in \eqref{dec}, followed by a spontaneous symmetry breaking of the dilaton field $\phi$ to account for the origin of Planck mass , would allow to recover the EH action from \eqref{cf}.

\subsection{Modified WZ subtractions: the scaleless $d=4$ anomaly effective action}

One may proceed  by introducing a finite renormalization/extension 
of the topological term, in order to derive a different version of $\sm_{WZ}$ \eqref{WZ}, which is quadratic in $\phi$, rather than quartic. This is obtained by extending the topological term at $O(\epsilon)$ in the form \cite{Mazur:2001aa}

\beq
E_{ext}=E_4 +\epsilon\frac{R^2}{2 (d-1)^2},
\label{eee}
\eeq
and the singular limit performed on the functional
\beq
\label{ss1}
\tilde{V}_E=\int d^d x \sqrt{g}E_{ext}. 
 \eeq
Also in this case $\sm_R$ can be defined by the inclusion of the modified action in replacement of the 
$\hat{V}'_E$ counterterm
\begin{equation}
\hat{V}'_E\equiv \mathcal{S}^{(WZ)}_{GB} \equiv\lim_{\epsilon\to 0}\frac{\alpha}{\epsilon}\left(\tilde{V}_E(\bar{g}_{\mu\nu}e^{2\phi},d)- \tilde{V}_E(\bar{g}_{\mu\nu},d\right).
\label{inter}
\eeq
This finite renormalizaton of the counterterm is limited to the Gauss-Bonnet term, while the expression of $\hat{V}'_{C^2}$ remain identical.  \\
To derive its nonlocal expression, we can use the relation
\beq
\frac{\delta}{\delta\phi}\frac{1}{\epsilon}\tilde{V}_E(g_{\mu\nu},d)= \sqrt{g}\left(E-\frac{2}{3}\Box R +
\epsilon\frac{R^2}{2(d-1)^2}\right)
\eeq
in \eqref{inter}, to obtain  
 \beqa
 \frac{\delta \mathcal{S}^{(WZ)}_{GB}}{\delta\phi}&=&\alpha\sqrt{g}\left(E-\frac{2}{3}\Box R \right)\nonumber \\
&=&\alpha\sqrt{\bar g}\left(\bar E-\frac{2}{3}\bar \Box\bar R + 4 \bar\Delta_4 \phi\right),
\label{solve}
\eeqa
and henceforth
\begin{equation}
\mathcal{S}^{(WZ)}_{GB} = \alpha\int\,d^4x\,\sqrt{-\bar g}\,\left\{\left(\overline E - {2\over 3}
\bar{\Box} \overline R\right)\phi + 2\,\phi\bar\Delta_4\phi\right\},\,
\label{WZ2}
\end{equation}

where $\Delta_4$ is the fourth order self-adjoint 
operator, which is conformal invariant when it acts on a scalar function of vanishing scaling dimensions 
\beq
\Delta_4 = \nabla^2 + 2\,R^{\mu\nu}\nabla_\mu\nabla_\nu - \frac23\,R{\Box}
+ \frac13\,(\nabla^\mu R)\nabla_\mu\,
\label{120}
\eeq
and satisfies the relation
\beq
\sqrt{-g}\,\D_4\chi_0=\sqrt{-\bar g}\,\bar{\D}_4 \chi_0,\label{point2}
\eeq
if $\chi_0$ is invariant (i.e. has scaling equal to zero) under a Weyl transformation. $\sm_{WZ}$ can be reformulated in a nonlocal form by eliminating $\phi$. For example, the vertex for three gravitons (three-wave interaction) extracted from this anomaly induced action, around flat space, is reproduced by the nonlocal expression

\begin{align}
&\hspace{-5mm} \mathcal{S}_{\rm anom}^{(3)} =
\sdfrac{b'}{9} \int\! d^4x \int\!d^4x'\!\int\!d^4x''\!\left\{\big(\partial_{\m} R^{(1)})_x\left(\sdfrac{1}{\sqb}\right)_{\!xx'}  
\!\left(R^{(1)\m\n}\! - \!\sdfrac{1}{3} \eta^{\m\n} R^{(1)}\right)_{x'}\!
\left(\sdfrac{1}{\sqb}\right)_{\!x'x''}\!\big(\partial_{\n} R^{(1)})_{x''}\right\}\notag\\
&\hspace{-5mm}- \sdfrac{1}{6} \int\! d^4x\! \int\!d^4x'\! \left(b'\, E^{\!(2)} + b\,  [C^2]^{(2)}\right)_{\!x}\! \left(\sdfrac{1}{\sqb}\right)_{\!xx'} \!R^{(1)}_{x'}
+ \sdfrac{b'}{18}  \int\! d^4x\, R^{(1)}\left(2\, R^{\!(2)} + (\sqrt{-g})^{(1)} R^{(1)}\right),
\label{S3anom3}
\end{align}
where the last term in the second line is purely local (i.e., no $\Box^{-1}$) and associated with the $\hat{V}'_{C^2}$ contribution. The vertex is organised in terms of metric fluctuations, with an expansion around flat spacetime up to second order.

\section{Chiral/gravitational anomalies in the early universe}
We now turn to discuss some salient features of the presence of chiral anomalies in the early universe, commenting on their  behaviour under the presence of chiral chemical potentials and thermal corrections.  Detailed analytical studies of these interactions using methods from conformal field theory, combined with perturbative approaches using quantum field theory at finite density, can be found in \cite{Coriano:2024nhv,Coriano:2023cvf,Coriano:2023gxa}. \\
Chiral anomalies play a critical role in the physics of the early universe, especially during periods when extreme conditions such as high temperatures, strong electromagnetic fields, and topologically non-trivial gauge fields were present. As shown in \eqref{rr}, as for conformal anomalies, these anomalies are quantum mechanical phenomena that arise when classical symmetries of a system, such as chiral symmetries, are broken at the quantum level.\\
Classically, the number of left-handed and right-handed fermions would be conserved separately in the absence of anomalies, if the underlying interactions are chirally symmetric. However, quantum effects, particularly the interaction of massless fermions with gauge fields (such as electromagnetic or gluonic fields), lead to what is known as a chiral anomaly, where this conservation law is violated.\\
In the early universe, particularly during the quark-gluon plasma (QGP) phase shortly after the Big Bang, the presence of strong electromagnetic fields and non-trivial gauge configurations (such as sphalerons in the electroweak sector) created an environment where these chiral anomalies could manifest and significantly impact the dynamics of the primordial plasma.
Spectral asymmetries caused by chiral anomalies—particularly the conventional chiral anomaly term \(F\tilde{F} \sim E \cdot B\)—have been studied for their influence on the evolution of the primordial plasma, impacting magneto-hydrodynamical (MHD) equations and the generation of cosmological magnetic fields \cite{Boyarsky:2011uy}.\\
As already mentioned, one of the hallmark expressions of the chiral anomaly is the non-conservation of the axial current in the presence of electromagnetic fields, described by the equation

\beq
\frac{d N_5}{dt} = \frac{e^2}{2\pi^2} \int d^3 x \vec{E} \cdot \vec{B},
\eeq
where \(N_5\) is the chiral fermion number (the difference between the number of right-handed and left-handed fermions), and \(\vec{E}\) and \(\vec{B}\) are the electric and magnetic fields. \\
One of the key effects associated with the chiral anomaly in the early universe is the Chiral Magnetic Effect (CME). The CME predicts that, in the presence of an imbalance between left-handed and right-handed fermions (chiral asymmetry) and a strong magnetic field, an electric current can be generated along the direction of the magnetic field
\beq
\vec{J} = \frac{e^2}{2\pi^2} \mu_5 \vec{B},
\eeq
where \( \mu_5 \equiv \mu_L-\mu_R\) is the chiral chemical potential, which quantifies the imbalance between left- and right-handed fermions.\\
In the early universe, strong magnetic fields were likely generated during phase transitions, such as the electroweak phase transition, or during processes like cosmic inflation. \\
Chiral anomalies are also closely related to the mechanism of baryogenesis and  the observed matter-antimatter asymmetry in the universe. 
In addition to the CME, the Chiral Vortical Effect (CVE) is another important consequence of chiral anomalies in the early universe. The CVE predicts that a rotating fluid with a chiral imbalance will generate a current along the direction of the fluid's vorticity. In the early universe, turbulence and rotational motions within the plasma, combined with chiral imbalances, could have led to the creation of additional currents and influenced the evolution of the plasma.\\
Therefore, the presence of strong magnetic fields and rotating fluids in the early universe enhances the complexity of the plasma's dynamics, with chiral anomalies playing a crucial role in shaping these processes.\\
Experiments involving heavy ion collisions have tested and confirmed the anomalous behavior of matter in the presence of finite density chiral asymmetric backgrounds and strong fields \cite{Huang:2015oca}. These experiments have provided crucial insights into the behavior of quark-gluon plasma (QGP), a state of matter that mimics the conditions present in the early universe, specifically within microseconds after the Big Bang, when the universe was extremely hot and dense. During this phase, the QGP behaves like a near-perfect fluid of deconfined quarks and gluons, and it is in this environment that the anomalous behavior of matter, governed by chiral asymmetry and strong electromagnetic fields, becomes prominent. \\
Thus, the study of the chiral anomaly and the behavior of matter under strong fields in heavy ion collisions not only reveals the fundamental properties of the QGP but also provides crucial connections to the plasma phase of the early universe. The anomalous behavior observed in modern experiments mirrors, in a laboratory setting, the conditions that existed during the infancy of the cosmos, giving us a deeper understanding of the physical processes that governed the evolution of the universe shortly after the Big Bang.
\subsection{Gravitational chiral anomaly}
The signature of anomalous interactions, being these conformal or chiral/gravitational is always associated with the exchange of a specific interaction, an anomaly pole, 
which differs kinematically by ordinary poles corresponding to ordinary particle states.
In free field theory, two realizations of currents have been studied in the context of such correlators: the bilinear (axial-vector) fermion current \(J_{5 f}\), and the bilinear gauge-dependent Chern-Simons (CS) current  \cite{Dolgov:1988qx, Dolgov:1987yp}. Both currents, generically denoted as $J_5$, can be expressed as either \eqref{five} or the Chern-Simons current
\beq
J_{CS}^{\lambda} = \epsilon^{\lambda \mu \nu \rho} V_{\mu} \partial_{\nu} V_{\rho}.
\eeq
where $V_\mu$ denotes a vector field.
The corresponding correlator is the $\langle J_5 TT \rangle$ 3-point function, where  $T$, as usual, denotes the stress-energy tensor.
Both $J_{5 f}$ and $J_{CS}$ can be considered in a perturbative realization of the same correlator and are capable of generating a gravitational anomaly. Notably, the second current, \(J_{CS}\), can be incorporated into a standard partition function—within a conventional Lagrangian formulation via a path integral—only in the presence of a coupling to an axial-vector gauge field \(A_{\lambda}\), through an interaction term given by
\beq
\mathcal{S}_{AVF} \equiv \int d^4 x \sqrt{g} A_\lambda J^\lambda_{CS}.
\eeq
This term, often denoted as \(A \, V\wedge F_V\), represents the Abelian Chern-Simons form. The Chern-Simons current \(J_{CS}\) mediates the gravitational chiral anomaly via virtual spin-1 photons in loops, leading to a disparity between their two circular polarizations and inducing optical helicity. Its integrated $0th$ component counts the difference in the number of photons of opposed helicities. 
This interaction is significant in early universe cosmology and affects the polarization of the Cosmic Microwave Background (CMB) \cite{Galaverni:2020xrq}. In other words, a trilinear interaction that couples to gravity via a correlator of the form $\langle J_{CS } TT\rangle $ where $T$ denotes the stress-energy tensor of matter and is contracted with the amplidue of a gravitational wave, may enhance the helicity asymmetry of spin-1 abelian fields.\\
In this context, the classical symmetry being broken could be the discrete duality invariance of the Maxwell equations in a vacuum, where \(E \to B\) and \(B \to -E\) (see \cite{Pasti:1995tn, Agullo:2018nfv}). Similarly, the \(\langle TTJ_5 \rangle\) correlator induces analogous effects on gravitational waves \cite{delRio:2020cmv, delRio:2021bnl}. As previously mentioned, our approach relies solely on the symmetries of the correlator to determine its structure, which remains valid for any generic parity-odd current \(J_5\). In both cases—whether involving \(J_{5f}\) or \(J_{CS}\)—the solution is fundamentally driven by the anomaly pole, which plays a crucial role in reconstructing the corresponding correlators.\\
It has been demonstrated in \cite{Dolgov:1988qx, Dolgov:1987yp} that both realizations of the \(\langle TTJ_5 \rangle\) correlator—whether involving a \(J_{CS}\) current or a \(J_{5f}\) current—reduce to the exchange of a special intermediate parity-odd state \cite{Giannotti:2008cv} \cite{Coriano:2014gja}.
 
\section{Conclusions}
Understanding the role of anomalies is crucial for explaining the transition from the highly symmetric, high-energy state of the early universe to the less symmetric, lower-energy state we observe today. In particular, the significance of the conformal anomaly in cosmological evolution is expected to be a key element in modern cosmology, as highlighted in \cite{Mavromatos:2022gtr}. The anomaly effective action exhibits an intriguing nonlocal structure, which is especially relevant for investigating the stochastic background of gravitational waves \cite{Mottola:2016mpl}. However, nonlocal cosmologies raise important questions that need to be addressed within the framework of the conformal anomaly-induced action \cite{Capozziello:2021bki}.\\
Regarding the chiral anomaly, during the early universe—specifically within the first few microseconds after the Big Bang—extreme temperatures and densities allowed strong interactions to dominate. The universe was filled with a hot plasma of quarks, gluons, leptons, photons, and other particles, where the effects of anomalies played a pivotal role. As the universe rapidly expanded and cooled, strong electric and magnetic fields would have formed, making anomalies and chiral effects, such as those described by Eq. (\ref{prima}), highly relevant to the dynamics of the primordial plasma. The non-conservation of chiral fermion number could have influenced critical processes like baryogenesis—the mechanism responsible for the matter-antimatter asymmetry in the universe—and contributed to the evolution of cosmic magnetic fields, which are observed on large scales today.

\vspace{0.3cm}
\centerline{\bf Acknowledgements}
This work is partially supported by INFN, inziativa specifica {\em QG-sky}, by the the grant PRIN 2022BP52A MUR "The Holographic Universe for all Lambdas" Lecce-Naples, and by the European Union, Next Generation EU, PNRR project "National Centre for HPC, Big Data and Quantum Computing", project code CN00000013. 


\providecommand{\href}[2]{#2}\begingroup\raggedright\endgroup

\end{document}